\documentclass[sigplan,screen]{acmart}

\copyrightyear{2026}
\acmYear{2026}
\setcopyright{cc}
\setcctype{by}
\acmConference[ASPLOS '26]{Proceedings of the 31st ACM International Conference on Architectural Support for Programming Languages and Operating Systems, Volume 2}{March 22--26, 2026}{Pittsburgh, PA, USA}
\acmBooktitle{Proceedings of the 31st ACM International Conference on Architectural Support for Programming Languages and Operating Systems, Volume 2 (ASPLOS '26), March 22--26, 2026, Pittsburgh, PA, USA}
\acmPrice{}
\acmDOI{10.1145/3779212.3790191}
\acmISBN{979-8-4007-2359-9/2026/03}
\usepackage{indentfirst} 
\usepackage{enumitem}
\usepackage{pifont}
\usepackage{ulem}
\usepackage{fontawesome5}
\usepackage{wrapfig}
\usepackage{graphicx,subcaption}
\usepackage{balance}
\usepackage{xcolor}
\usepackage{listings}
\usepackage{comment}

\definecolor{darkgreen}{rgb}{0.0,0.5,0.0}

\newcommand{\ZNSKVCache}{Nemo}

\title{\ZNSKVCache: A Low-Write-Amplification Cache for Tiny Objects on Log-Structured Flash Devices}

\author{Xufeng Yang}
\affiliation{%
  \institution{Xiamen University}
  \city{Xiamen}
  \country{China}}
\email{xufengyang@stu.xmu.edu.cn}

\author{Tingting Tan}
\affiliation{%
  \institution{Xiamen University}
  \city{Xiamen}
  \country{China}}
\email{tingtingtan@stu.xmu.edu.cn}

\author{Jingxin Hu}
\affiliation{%
  \institution{Chongqing University of Posts and Telecommunications}
  \city{Chongqing}
  \country{China}}
\email{hugeorcyx@gmail.com}

\author{Congming Gao\footnotemark}
\affiliation{%
  \institution{Xiamen University}
  \city{Xiamen}
  \country{China}}
\email{gaocm92@gmail.com}

\author{Mingyang Liu}
\affiliation{%
  \institution{Xiamen University}
  \city{Xiamen}
  \country{China}}
\email{mingyang@stu.xmu.edu.cn}

\author{Tianyang Jiang}
\affiliation{%
  \institution{Openharmony Community}
  \city{Beijing}
  \country{China}}
\email{jty1313@126.com}

\author{Jian Chen}
\affiliation{%
  \institution{Tsinghua University}
  \city{Beijing}
  \country{China}}
\email{jian-che21@mails.tsinghua.edu.cn}

\author{Linbo Long}
\affiliation{%
  \institution{Chongqing University of Posts and Telecommunications}
  \city{Chongqing}
  \country{China}}
\email{longlb@cqupt.edu.cn}

\author{Yina Lv}
\affiliation{%
  \institution{Xiamen University}
  \city{Xiamen}
  \country{China}}
\email{lvyina@xmu.edu.cn}

\author{Jiwu Shu}
\affiliation{%
  \institution{Xiamen University}
  \city{Xiamen}
  \country{China}}
\affiliation{%
  \institution{Tsinghua University}
  \city{Beijing}
  \country{China}}
\email{jwshu@xmu.edu.cn}

\begin{document}

\begin{abstract}
Modern storage systems predominantly use flash-based SSDs as a cache layer due to their favorable performance and cost efficiency. 
However, in tiny-object workloads, existing flash cache designs still suffer from high write amplification.
Even when deploying advanced log-structured flash devices (e.g., Zoned Namespace SSDs and Flexible Data Placement SSDs) with low device-level write amplification, application-level write amplification still dominates.

This work proposes \ZNSKVCache, which enhances set-associative cache design by increasing hash collision probability to improve set fill rate, thereby reducing application-level write amplification. 
To satisfy caching requirements, including high memory efficiency and low miss ratio, we introduce a bloom filter-based indexing mechanism that significantly reduces memory overhead, and adopt a hybrid hotness tracking to achieve low miss ratio without losing memory efficiency. 
Experimental results show that \ZNSKVCache~ simultaneously achieves three key objectives for flash cache: low write amplification, high memory efficiency, and low miss ratio.
\end{abstract}



\maketitle

\footnotetext{*Corresponding author}

\section{Introduction}
\label{sec:intro}
Key–value (KV) caches are widely employed in modern data centers to accelerate a broad range of services, including KV stores, social media applications, and content delivery networks \cite{zhu2024memory, song2023prism, shen2023fusee, tiktok, Rocksdb, Leveldb, chen2020flatstore, Redis, memcached, couchbase}. These services often manage massive volumes of tiny objects, each only a few hundred bytes or less \cite{mcallister2021kangaroo}.
As these services continue to grow in scale, the demand for larger cache capacity has increased significantly. 
Therefore, the high cost and limited scalability of memory-based caches has prompted a shift toward flash-based design that leverages SSDs as the cache media for improved cost-efficiency and capacity \cite{tang2015ripq, shen2018didacache, fleisenman2019flashield, berg2020cachelib, mcallister2021kangaroo, yang2022cachesack, mcallister2024fairywren, yang2024can, chen2024space, shu2024data}. 

Flash-based KV cache faces a new set of challenges that do not arise in traditional memory-based cache. 
Unlike memory, the inherent properties of flash memory, such as out-of-place updates and limited write endurance, fundamentally reshape design priorities \cite{gao2019constructing, micheloni2013inside, lu2014reconfs}.
Therefore, in addition to classic  metrics like miss ratio and serving latency, flash cache must also manage memory overhead coming from indexing structures and mitigate write amplification (WA) at both application-level WA (ALWA) and device-level WA (DLWA) \cite{mcallister2021kangaroo, mcallister2024fairywren, allison2025towardsFDPcache}. 
These two metrics are particularly critical under tiny-object workloads: the number of index entries per unit of flash increases sharply, raising memory overhead; meanwhile, the mismatch between object size and flash write granularity leads to significant write amplification, accelerating device wear.
As a result, designing an efficient flash-based KV cache requires jointly optimizing caching performance, memory overhead, and flash friendliness.

\begin{figure}[t]
    \centering
    \includegraphics[width=0.97\linewidth]{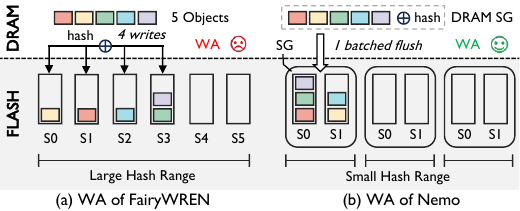}
    \vspace{-0.15in}
    \caption{Application-level write amplification comparison. ``S'' denotes Set,  ``SG'' denotes Set-Group.}
    \label{fig:alwa-fw-nemo}
     \vspace{-0.26in}
\end{figure}

Existing flash cache designs have achieved memory cost optimization for tiny-object workloads, yet they still exhibit high WA \cite{mcallister2021kangaroo,mcallister2024fairywren}.
To address this, several SSD manufacturers, such as Western Digital and Samsung, have introduced hardware-level solutions such as Zoned Namespace (ZNS) SSDs \cite{bjorling2021zns} and Flexible Data Placement (FDP) SSDs \cite{samsungFDPintroduction}.
These advanced log-structured SSDs integrate application-managed data placement with the underlying flash-friendly write patterns, thereby minimizing DLWA.
However, the ALWA of log-structured SSD based cache remains substantial. 
For instance, FairyWREN \cite{mcallister2024fairywren}, a state-of-the-art (SOTA) solution for tiny objects, exhibits an ALWA exceeding $\textbf{15}\times$ when processing Twitter traces (average object size: 246 bytes) \cite{yang2021large, twitter-trace} on a commercial ZNS SSD (ZN540 \cite{ZN540}). 
In this work, to identify the root cause of high ALWA and establish design principles for our work, we performed theoretical analysis and practical evaluation (§\ref{sec:motivation}).
The analysis and evaluation results show that FairyWREN's high ALWA stems from the set-associative mapping with a large hash space (as illustrated in Figure~\ref{fig:alwa-fw-nemo}(a)). Specifically, flushing only a small number of tiny objects into a single set at a time results in inefficient write batching (i.e., a low set fill rate $\approx$7.0\% based on our evaluation), contributing $\approx$14.20 of the total ALWA.

To address this, we propose \ZNSKVCache, a new flash cache architecture designed for tiny objects that \textbf{achieves near-ideal write amplification without increasing memory cost and caching performance.}
As shown in Figure~\ref{fig:alwa-fw-nemo}(b), \ZNSKVCache~ employs a set-associative mapping with a small hash space (i.e., a small in-memory Set-Group, SG), which contains multiple fixed-size sets (e.g., S0 and S1 in this figure), to improve the set fill rate.
In practice, while hashed keys follow a uniform distribution in the long run, they exhibit noticeable skew during the short period when an SG transitions from being empty to having any set filled.
Therefore, \ZNSKVCache~ smooths out this skew with three mechanisms (\S~\ref{sec:perfect_SG}), boosting the average set fill rate from 7.0\% to \textbf{over 89\%}.
These high-fill-rate SGs are then flushed to SSD and organized into a FIFO-managed on-flash SG pool.
When cache space becomes insufficient, \ZNSKVCache~ performs SG-level eviction. 
This cache flush and eviction pattern aligns with SSD's log-structuring feature through its batched write and invalidation behavior. 
As a result, \ZNSKVCache~ achieves an ALWA of 1.56 close to the theoretical optimum, and its DLWA can be as low as 1 on existing log-structured SSDs (e.g., ZNS, FDP, and conventional namespace SSDs). 

With the low write amplification, \ZNSKVCache~ also maintains a low memory cost for metadata, including object indexing and hotness tracking. 
First, \ZNSKVCache~ reduces indexing overhead by approximating object indexing. 
Specifically, we replace exact per-object mapping with a Parallel Bloom Filter Group (PBFG) composed of multiple bloom filters.
This enables parallel queries across filters for object location identification while capitalizing on the space efficiency of bloom filters. 
Second, \ZNSKVCache~ reduces memory usage by selectively offloading metadata to flash based on update frequency. 
Frequently updated metadata, such as object hotness, are kept in memory, whereas mostly stable metadata, such as object indexes (e.g., PBFG), are offloaded to flash, with only frequently accessed indexes cached in memory.
Third, \ZNSKVCache~ employs a hybrid mechanism to track object hotness with minimal memory overhead.
Specifically, with the hotness of in-memory indexes, \ZNSKVCache~ implements hotness tracking as a coarse-grained, 1-bit-per-object, avoiding the need for fine-grained and memory-intensive access counters.

We implement \ZNSKVCache~ as a module in CacheLib \cite{berg2020cachelib} and evaluate it by replaying Twitter traces on a real log-structured flash device.
Results show that \ZNSKVCache~ achieves low WA without sacrificing memory efficiency and performance.
The main contributions are as follows:
\begin{itemize} [leftmargin=12pt, itemsep=0pt, topsep=0pt]
\item We construct an model to characterize and explain the sources of write amplification in FairyWREN (\S \ref{sec:motivation}).
\item We propose \ZNSKVCache, a new and general flash caching framework, which rearchitects set-associative caching to reduce write amplification and leverages an on-demand approximate index to keep memory usage low (\S \ref{sec:design}).
\item Results show that \ZNSKVCache~ cuts flash writes by up to 90\% with only 8.3 bits of memory per object (\S \ref{sec:eval}).

\end{itemize}

\section{Background}
\label{sec:bg}
\subsection{Tiny objects in KV cache}
KV systems need to store massive numbers of tiny objects, each only a few hundred bytes or less \cite{mcallister2021kangaroo, berg2020cachelib, yang2021large, bronson2013tao, li2017workload, li2018better}. 
These systems typically handle millions of object accesses.
For example, TikTok generates nearly 575.6 million new comments daily, each capped at 200 B \cite{tiktok_data1, tiktok_data0}.
Twitter logs over 500 million new tweets limited to 280 B per one \cite{twitter_data0, twitter_data1}.
These tiny objects are typically managed by KV store \cite{ousterhout2015ramcloud, Leveldb, Rocksdb, SkimpyStash, SpanDB, zoneDB, lepers2019kvell}.
To enhance performance, a caching layer referred to as a KV cache is integrated into the storage hierarchy \cite{berg2020cachelib, memcached, Redis, couchbase}. 
However, KV cache is not simply a condensed version of KV store, although they expose similar APIs. 
The key distinction is that: deletion is user-driven, while eviction is initiated by the cache when space is exhausted. Therefore, KV caches have greater flexibility in deciding what to store.

\subsection{The design criteria of flash cache}
\label{sec:main_metrics}
In-memory KV cache is fundamental to modern systems, and its effectiveness is reflected by two key metrics: \underline{\textit{\textbf{hit ratio}}} and \underline{\textit{\textbf{serving performance}}} 
\cite{TTLmatters, GL-Cache, segcache, metacacheanalysis, mcallister2021kangaroo, berg2020cachelib, 3L-cache, zhang2017flashkv, nan2025ACCACHE}.
The hit ratio, defined as the fraction of requests served from memory, directly determines the backend load and is largely governed by the cache size and eviction policy.
Serving performance, characterized by tail latency and/or throughput, is crucial for meeting workload demands. 
As working set sizes continue to grow, scaling memory-based cache to sustain both high hit ratio and low latency has become increasingly cost-prohibitive \cite{tang2015ripq, berg2020cachelib}.
In contrast, modern SSDs have made the cost-per-bit more than an order of magnitude lower than that of memory, while they deliver bandwidths in the multi-GB/s range \cite{DRAMeXchange, bestSSD}.
Consequently, modern systems adopt flash-based caches \cite{tang2015ripq, berger2017adaptsize, li2017pannier, shen2018didacache, fleisenman2019flashield, berg2020cachelib, mcallister2021kangaroo, mcallister2024fairywren, chen2024space}. 
This architectural shift introduces additional evaluation dimensions, including memory overhead and write amplification.

\uline{\textit{\textbf{Memory overhead:}}} In flash-based caches, objects are stored on flash with a small amount of memory used for metadata. 
As flash is introduced to reduce the reliance on memory, excessive memory overhead can undermine this goal. 
This tension forces flash cache design to trade indexing efficiency and performance against memory footprint carefully.

\uline{\textit{\textbf{Write amplification (WA):}}}
Due to the limited endurance of flash memory \cite{boboila2010write, luo2015warm, flashCell, yi2024biza, gao2022reprogramming}, flash caches must carefully manage WA. 
In general, the overall WA originates from two sources: application-level write amplification (ALWA) and device-level write amplification (DLWA).

\textbf{ALWA} occurs when user data is rewritten due to the data management mechanisms. Specifically, first, fine-grained writes whose size is smaller than a logical block introduces read–modify–write (RMW) overhead, causing write amplification. 
A common mitigation is to buffer and batch these small writes \cite{shrira1994opportunistic}. 
Second, the append-only updates, such as log-structured design \cite{Leveldb,Rocksdb,f2fs,segcache,lu2013extending}, require the obsolete or invalid data to be reclaimed through cleaning or compaction, causing valid data to be rewritten.
This overhead is typically mitigated by exploiting update locality \cite{he2017unwritten, dubeyko2019ssdfslfsflashfriendlyfile, zoneKV, sun2022light, li2025bvlsmwriteefficientlsmtreestorage}.

\textbf{DLWA} stems from device-internal flash management mechanisms \cite{micheloni2013inside, hardock2017place}.
NAND flash enforces that data must be erased at block granularity before any page within can be rewritten.
This constraint requires SSD to perform garbage collection, which relocates valid pages and incurs extra writes \cite{agrawal2008design, lee2011semi, 202247, li2021ioda, yang2024place}.
A common solution is to increase over-provisioning (OP) space, exposing only part of the raw capacity to the host. This reduces the fraction of valid pages in each block and thereby lowers garbage collection overhead. 
Alternatively, DLWA can be reduced by shaping application write patterns to issue large, sequential writes, or by leveraging advanced device interfaces such as ZNS and FDP \cite{NVMExpress-Specification, OCSSD-2017lightnvm, 2017enhancingSSD-multiS, yang2017autostream, bjorling2021zns, tehrany2022understandingZNS, allison2025towardsFDPcache}.
These interfaces allow the host to issue data placement directives and lifetime hints, enabling the device to group data with similar lifetimes.
As a result, pages within a block tend to be invalidated together, allowing blocks to be reclaimed with minimal data movement and reduced DLWA.

\begin{table}[t]
\setlength{\tabcolsep}{1.5mm}{
\caption{Comparison of prior cache designs (with \ZNSKVCache).}
\label{tab:compare}
 \vspace{-0.12in} 
\small
\begin{tabular}{cccc}
\hline
                       & \textit{\textbf{Write Amp.}} & \textit{\textbf{Memory}} & \textit{\textbf{Flash Util.}} \\ \hline
Log-structured          &              {\color{darkgreen}\ding{52}\ding{52}}                         &       {\color{purple}\ding{56}\ding{56}}                                                         &     {\color{darkgreen}\ding{52}\ding{52}}                                \\
Set-associative        &        {\color{purple}\ding{56}\ding{56}}        &          {\color{darkgreen}\ding{52}\ding{52}}                       &          {\color{purple}\ding{56}\ding{56}}            \\
Hierarchical              &           {\color{purple}\ding{56}}                             &            {\color{darkgreen}\ding{52}}                               &         {\color{darkgreen}\ding{52}}                            \\
\textit{\textbf{\ZNSKVCache}} &           {\color{darkgreen}\ding{52}\ding{52}}                             &        {\color{darkgreen}\ding{52}}                      &           {\color{darkgreen}\ding{52}\ding{52}}                           \\ \hline
\end{tabular}}
  \vspace{-0.23in}
\end{table}

\subsection{Limitations of Existing Approaches}
\label{sec:limitation}
Table~\ref{tab:compare} summarizes three typical features of flash cache.
The first adopts a \textbf{log-structured} design to minimize flash writes \cite{tang2015ripq, berger2017adaptsize, li2017pannier, shen2018didacache, fleisenman2019flashield}. 
Objects are buffered in memory and then appended to a flash log in batches, resulting in ideal write amplification.
However, for tiny objects, the in-memory index becomes expensive. 
Each entry stores a flash offset ($\sim$29 bits), a tag ($\sim$29 bits), a next pointer (64 bits), and optional hotness, totaling over 15 B per object. In tiny object scenarios, this overhead can be as high as 10\% of the object size, making log-structured designs expensive at scale.

The second type, \textbf{set-associative} design, as implemented by Meta (CacheLib \cite{berg2020cachelib}), reduces memory overhead for tiny objects by hashing keys into fixed sets.
Each set corresponds to a 4 KB flash page, and objects are read at page granularity, so per-object flash address mapping is not required. 
However, this approach makes flash prone to excessive writes. 
Inserting a tiny object (e.g., 200 bytes) requires rewriting the entire 4 KB set, yielding an ALWA of up to 20$\times$.
Within the device, random object placement produces scattered invalid pages, forcing garbage collection and increasing DLWA. 
To eliminate DLWA, Meta adopts 50\% OP in production deployments, which results in low flash utilization \cite{berg2020cachelib}.

Log-structured and set-associative caches represent two extremes in the design space, exposing a fundamental trade-off between write amplification and memory overhead. The
\begin{wrapfigure}{l}{0.48\columnwidth}
  \centering
  \includegraphics[width=0.52\columnwidth]{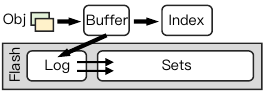}
   \vspace{-0.32in}
  \caption{Hierarchical cache.}
  \vspace{-0.23in}
  \label{fig:hc}
\end{wrapfigure}
third type, \textbf{hierarchical design} (i.e., Kangaroo \cite{mcallister2021kangaroo} and FairyWREN \cite{mcallister2024fairywren}) has emerged as a SOTA solution, combining a log-structured front tier with a set-associative back tier to balance write amplification and memory cost.
As shown in Figure \ref{fig:hc}, the front tier (i.e., Hierarchical-Log, HLog) buffers a large number of objects and then migrates a small group of objects that map to the same set into the back tier (i.e., Hierarchical-Set, HSet).
This design allows each HSet flash write to insert multiple objects, reducing write amplification.
To enable this in practice, HLog maintains a hash table with linked-list–based collision resolution, ensuring the table entries number equals the number of sets in HSet. 
Each entry maintains a linked list that records all objects mapped to the same set.
However, existing hierarchical caches still incur substantial write amplification under tiny-object workloads.

\section{Motivation: Miles to Go in WA Reduction}
\label{sec:motivation}
Kangaroo and FairyWREN are two known hierarchical cache designs, with FairyWREN building on Kangaroo to optimize write amplification. 
FairyWREN employs a host-managed mapping layer to merge garbage collection and log-to-set migration into a single operation, thereby reducing flash writes.
Consequently, unlike Kangaroo, which treats ALWA and DLWA separately, FairyWREN integrates DLWA into ALWA. 
In this work, we define Kangaroo's WA as the product of ALWA and device-level garbage collection overhead, while for FairyWREN, we report only ALWA.

As FairyWREN is the successor to Kangaroo, we reproduced the Twitter trace using its open-source implementation \cite{FW-sourcecode}. During this process, we discovered two major implementation bugs in the HLog-to-HSet migration logic, along with several minor issues such as compile errors and incorrect statistics. The major bugs caused most objects to never reach the HSet, meaning that only writes to HLog were reflected in the measured write amplification, while HSet's contribution was significantly underreported. 
As a result, the write amplification reported in the original paper was lower than the actual value.
After fixing these issues, we observed that FairyWREN exhibits a write amplification exceeding 15$\times$ under the same workload. 

To understand the cause of this high write amplification, we conducted a comprehensive case analysis.
\S\ref{sec:alwa_breakdown} identifies the major sources, and \S\ref{sec:moti_quantify_L2SWA} models and validates them using real-world traces and experiments on real devices.

\subsection{Breakdown of the WA in hierarchical cache}
\label{sec:alwa_breakdown}

\begin{figure}
    \centering
    \includegraphics[width=0.97\linewidth]{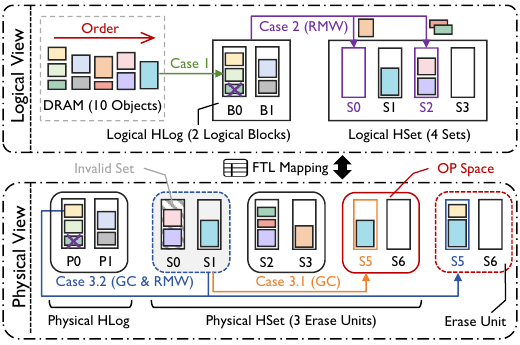}
     \vspace{-0.15in}
    \caption{
    Breakdown of write amplification in a hierarchical cache. ``P0$\sim$P1'' denotes 2 flash pages in the physical HLog. ``S5$\sim$S6'' appears twice, but there is only one physical instance, to illustrate the differing garbage collection triggering behaviors of FairyWREN (Case 3.2) and Kangaroo (Case 3.1).} 
    \label{fig:alwa-fw}
     \vspace{-0.22in}
\end{figure}

Figure~\ref{fig:alwa-fw} illustrates three sources of write amplification in hierarchical caches using an example with 10 inserted objects.
Case~1 occurs when objects are flushed from memory to the front tier, and Case~2 occurs when the front tier log is full and objects are migrated to back tier sets.
Case~$3.x$ is triggered by garbage collection in the hierarchical cache (Case~3.1 for Kangaroo and Case~3.2 for FairyWREN). We analyze the write path from two perspectives. From the logical view, we examine Case~1 and Case~2, which are independent of flash-level behavior. From the physical view, we analyze Case~$3.x$, as it involves interactions with flash-level garbage collection.

In \textbf{Case 1}, the write amplification stems from the size of objects not being aligned with 4 KB of flash page. 
Let the page fill rate during each flush be denoted as \( FR_i \).
The write amplification of Case 1 can be expressed as \( \frac{1}{\mathbb{E}(FR_i)} \), where $\mathbb{E}(FR_i)$ denotes the expected value of per-page fill rate.

In \textbf{Case 2}, objects are migrated from HLog to HSet, where the target sets are either written for the first time or updated. The former occurs only during the early stage of cache operation, whereas the latter dominates once the cache reaches steady state and incurs ``read-modify-write'' (RMW) operations, along with updates to FTL mapping.
In both cases, the resulting write amplification equals the set size divided by the expected value of the total size of newly written objects to the set.
We refer to this type of write amplification as \textbf{Log-to-Set Write Amplification (L2SWA)}.

When the flash cache is full, the hierarchical cache triggers garbage collection to reclaim invalid pages produced by earlier RMW operations. At this point (\textbf{Case 3.1}), write amplification comes from migrating valid sets within the reclaimed erase unit. We refer to this as \textbf{Garbage Collection Write Amplification (GCWA)}. While the cache could choose to drop these valid sets instead of migrating them, this is not recommended: these sets are carefully constructed, containing hot objects. Discarding these objects would lead to repeated RMW operations to repopulate the sets and a noticeable drop in cache utilization.

GCWA is highly sensitive to the OP ratio of HSet: a larger OP space eases garbage collection but reduces effective cache capacity. 
In practice, hierarchical caches often operate under tight OP constraints (e.g., 5\%), which significantly exacerbates GCWA.
To mitigate GCWA, FairyWREN integrates garbage collection with log-to-set migration by leveraging host FTL (\textbf{Case 3.2}). 
As shown in the physical view of Figure~\ref{fig:alwa-fw}, when reclaiming valid pages from an evicted erase unit, FairyWREN retrieves objects from HLog whose target Set IDs match those valid pages, and writes them together (indicated by the dark blue line). This way, these objects in HLog are migrated to HSet, skipping Case 2. The process resembles a variant RMW operation: it reads two pages, one from HSet and one from HLog, and writes one in HSet. As a result, in FairyWREN, GCWA is folded into L2SWA, making it more efficient than traditional garbage collection like Kangaroo, which follow Case 3.1.

In conclusion, FairyWREN incurs fewer flash writes than Kangaroo. Its write amplification can be expressed as:
\begin{equation}
  \text{\textbf{WA(FairyWREN)}} = \frac{1}{\mathbb{E}(FR_i)} +
\text{\textbf{L2SWA}} 
\label{eq:1}
\end{equation}

In the tiny-object workload, each appended page from the memory cache achieves a high fill rate, indicating that the first term in the formula is close to 1. 
However, prior measurements show that FairyWREN's overall write amplification can still reach up to 15$\times$, implying that \textbf{L2SWA dominates the total write amplification}. We now analyze the underlying reasons.


\begin{figure*}[t]
  \centering
  \begin{minipage}[b]{0.33\textwidth}
    \includegraphics[width=0.99\linewidth]{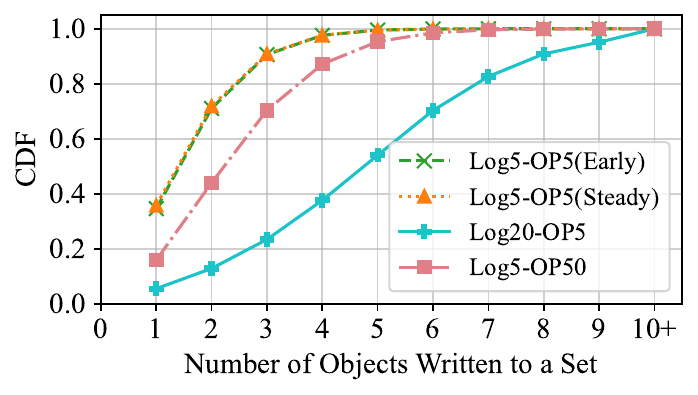}
\vspace{-0.15in}
    \caption{Passive object migration.}
    \label{fig:moti_ob1_ob2}
  \end{minipage}
  \hfill
  \begin{minipage}[b]{0.33\textwidth}
    \includegraphics[width=0.99\linewidth]{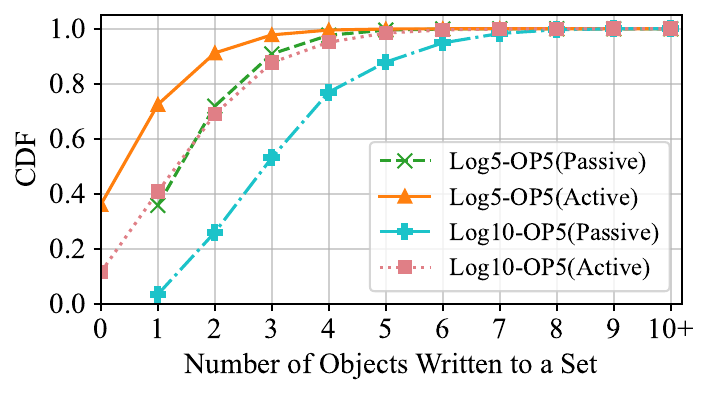}
\vspace{-0.15in}
    \caption{CDF of two migrations.}
    \label{fig:moti_ob3}
  \end{minipage}
  \hfill
  \begin{minipage}[b]{0.33\textwidth}
    \includegraphics[width=0.98\linewidth]{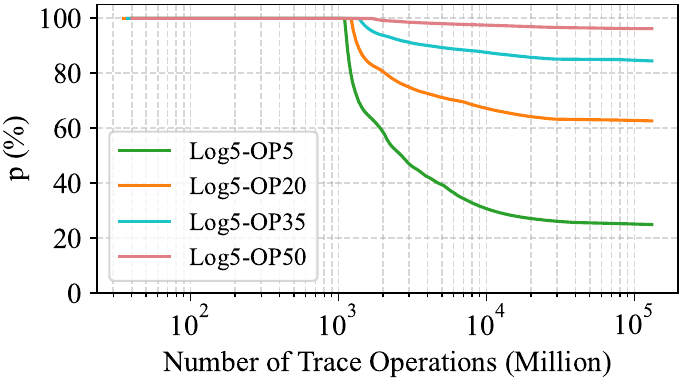}
\vspace{-0.15in}
    \caption{OP impact on passive migration.}
    \label{fig:moti_ob4}
  \end{minipage}
  
\vspace{-0.1in}
\end{figure*}

\subsection{Quantifying the Impact of L2SWA}
\label{sec:moti_quantify_L2SWA}

L2SWA is contributed by Cases 2 and 3\footnote{In this subsection, Case 3.2 is referred to simply as Case 3.}. During the early stage of cache running, it mainly stems from Case 2, while in the later stage, it arises from Case 2 and 3.
Meanwhile, Case 3 differs from Case 2 by prematurely migrating objects in HLog, leading to distinct contributions to L2SWA. Thus, we define the WA from Case 2 (\uline{\textbf{p}}assive object migration) as \textbf{L2SWA (P)} and that from Case 3 (\uline{\textbf{a}}ctive object migration) as \textbf{L2SWA (A)}. Let \(p\ (p<1)\) denote the proportion of RMW operations from Case 2, with (\(1-p\)) from Case 3. Then: 
\begin{equation}
\text{\textbf{L2SWA}} = 
p \cdot \text{\textbf{L2SWA(P)}} + (1 - p) \cdot \text{\textbf{L2SWA(A)}}
\label{eq:2}
\end{equation}

As both \textbf{L2SWA(P)} and \textbf{L2SWA(A)} correspond to RMW operations,
the log-to-set write amplification can be expressed as the ratio between
the set size and the expected value of the total size of newly written objects per set write:
\begin{equation}
    \text{\textbf{L2SWA(}}\ast\text{\textbf{)}} \triangleq
\frac{\text{Set size}}{\mathbb{E}(\text{Newly object total size per set write})}
\label{eq:3}
\end{equation}
This expected total size is related to the expected length of the linked list in the HLog hash table, denoted as \(\mathbb{E}(L_i)\).

Before the following analysis, we introduce the basic variables; the notations are summarized in Table~\ref{tab:keyvars-l2swa}.
We first state that the size of a set is identical to the flash page, and is denoted as \(w\). Then, let \(N_{\text{Log}}\) and \(N_{\text{Set}}\) denote the total numbers of pages in HLog and HSet, respectively. Note that \(N_{\text{Set}}\) does not equal the number of usable sets in HSet, since part of the space must be reserved as OP to support garbage collection. As a result, the actual number of sets, denoted as \(N'_{\text{set}}\), is smaller than \(N_{\text{Set}}\).
To characterize \(N'_{\text{set}}\), we redefine the OP ratio in a simplified but equivalent form. Rather than adopting the conventional definition based on the ratio of physical capacity minus user capacity to user capacity, we let the OP ratio \(X\)  denote the fraction of HSet reserved for garbage collection. Consequently, only the remaining fraction \((1-X)\) of HSet is usable for caching. That is,
\begin{equation}
N'_{\text{set}} = (1 - X)\, N_{\text{Set}}
\label{eq:4}
\end{equation}

\begin{table}[t]
\centering
\small
\caption{Key variables used in \S3.2.}
\vspace{-0.1in}
\begin{tabular}{cl}
\toprule
\textbf{Symbol} & \textbf{Description} \\
\midrule
$L2SWA$ & Overall log-to-set write amplification \\
$L2SWA(P)$ & L2SWA under Case~2 (passive migration) \\
$L2SWA(A)$ & L2SWA under Case~3 (active migration) \\
\midrule
$p$ & Fraction of RMWs that fall into Case~2 \\
w & Set (page) size \\
$N_{\mathrm{Log}}$ & Number of flash pages in HLog \\
$N_{\mathrm{Set}}$ & Number of flash pages in HSet \\
$N'_{\mathrm{Set}}$ & Number of sets in HSet \\
$X$ & Over-provisioning ratio of HSet \\
\bottomrule
\end{tabular}
\label{tab:keyvars-l2swa}
\vspace{-0.05in}
\end{table}

\subsubsection{Analysis of L2SWA(P)}

We assume that the expected object size is \(s\).
The maximum number of objects per page is \(\frac{w}{s}\), thus HLog can hold \(\frac{w}{s} \cdot N_{\text{Log}}\) objects.
The HLog's hash table contains \(\frac{1}{2}N'_{\text{Set}}\) linked list, corresponding to the cold set number in HSet due to FairyWREN's hot-cold division. 
Over time, the hash function ensures a near-uniform distribution, leading to balanced list lengths with minimal skew. Then, the expected length of each linked list \(\mathbb{E}(L_i)\) is:
\begin{equation}
 \mathbb{E}(L_i) = \frac{\frac{w}{s}\cdot N_{\text{Log}}}{\frac{1}{2} N'_{\text{Set}}} = \frac{2w\, N_{\text{Log}}}{s\, N'_{\text{Set}}}  
 \label{eq:5}
\end{equation}
Since a set write flushes all objects from a HLog linked list, the expected size of the newly set write is 
$\mathbb{E}(L_i)\cdot s$.
Therefore, according to the definition of log-to-set write amplification, L2SWA(P) is derived from Equation \eqref{eq:3}, \eqref{eq:4} and \eqref{eq:5}:
\begin{equation}
    \text{\textbf{L2SWA(P)}} = \frac{w}{\mathbb{E}(L_i) \cdot s}
= \frac{N'_{\text{Set}}}{2N_{\text{Log}}}
= \frac{(1-X)\, N_{\text{Set}}}{2N_{\text{Log}}}
\label{eq:6}
\end{equation}

\textbf{\textit{Theory vs. Practice:}} 
To investigate the practical \(\text{L2SWA(P)}\) of FairyWREN, we evaluated it on a real ZNS SSD using Twitter traces (average 246 B objects). The workload was run long enough to trigger garbage collection and active object migration, allowing us to examine its impact on passive object migration. 
Figure \ref{fig:moti_ob1_ob2} shows the cumulative distribution of newly written objects per set write under different configurations. 
``Early'' refers to the period before the cache first triggers garbage collection, while ``Steady'' denotes the long-running phase in which two migration behaviors have stabilized.
The results show that the distribution of passive object migration is nearly identical in the early and steady states.
With a Log/Set ratio of 5\%/95\% and 5\% OP, 71\% of write batches contain no more than 3 objects and 91\% contain no more than 4, indicating that only a small number of objects are newly written in each set write. And the measured \(\text{L2SWA(P)}\) is 8.5, which closely matches the theoretical value ($\approx$9) calculated from Equation \eqref{eq:6}. 
This shows that \uline{\(\text{L2SWA(P)}\) is substantial and depends only on the number of flash pages in HLog and the number of sets in HSet (i.e., the hash range of set-associative design), and independent of active object migration. (\textbf{Observation 1})}

We further conducted two experiments, also shown in Figure \ref{fig:moti_ob1_ob2}. First, we increased HLog to 20\% of the total flash space, incurring about 72\% more memory. 
Second, we increased the HSet OP ratio to 50\%, leaving half of the flash space unused. 
In both cases, the number of newly written objects per set write increases but remains limited: fewer than 20\% of set writes have more than 7 newly written objects in the former, and more than 4 in the latter. This shows that \uline{increasing HLog and/or reducing the number of sets in HSet can slightly mitigate \(\text{L2SWA(P)}\), but with high memory cost and reduced flash utilization. (\textbf{Observation 2})}

\subsubsection{Analysis of L2SWA(A)}

To quantify active migration's impact, we analyze the expected value of objects' lifespan (\(L\)) in HLog.
Assume that under passive migration (Case 2), an object remains in HLog for \(L(P) = T_i\). 
Under the uniform-hash assumption, the residence time of an actively migrated object (Case 3) is approximately uniformly distributed over \([0, T_i]\), i.e., (\(L \sim \mathcal{U}(0, T_i)\)). 
The expected lifespan of actively migrated objects is then \(\mathbb{E}(L(A)) = \frac{T_i}{2}\). 
The shorter residence time reduces the number of objects associated with each hash bucket at any given moment.
This leads to \(\text{L2SWA(A)} = 2 \cdot \text{L2SWA(P)}\).  
Therefore, \(\text{L2SWA}\) in Equation \eqref{eq:2} can be simplified to: 
\begin{equation}
  \text{\textbf{L2SWA}} = (2 - p) \cdot \text{\textbf{L2SWA(P)}}  
  \label{eq:7}
\end{equation}

\textbf{\textit{Theory vs. Practice:}} We first evaluated FairyWREN's $\text{L2SWA}$ under a Log/Set ratio of 5\%/95\% with 5\% OP. The observed \(\text{L2SWA}\) is 14.2 and \(p\) is stabilized around 25\% (detailed in \ref{sec:analysisP}), closely matching the theoretical value of 15.75 derived from the expression \((2-p)\cdot 9\). We further analyzed the cumulative distribution of the number of objects newly written to a set under passive and active object migration (Figure~\ref{fig:moti_ob3}). Under the same configuration, each set averagely receives 2.04 new objects per write during passive migration and 1.03 during active migration, consistent with the observed \(2\times\) gap between \(\text{L2SWA(P)}\) and \(\text{L2SWA(A)}\).
The same trend is observed with a Log/Set ratio of 10\%/90\% and 5\% OP.
These prove that \uline{active object migration incurs higher write amplification than passive object migration, with \(\text{L2SWA(A)}\) roughly twice \(\text{L2SWA(P)}\). (\textbf{Observation 3})}

\subsubsection{Analysis of \textit{p}}
\label{sec:analysisP}
The proportion of RMW operations caused by passive object migration is strongly correlated with the OP ratio: a higher OP ratio leads to a higher \(p\).
We conduct a controlled experiment to evaluate the impact of different OP ratios (Figure~\ref{fig:moti_ob4}). 
The value of \(p\) initially equals 100\% because the HSet is empty. As active object migration begins, its proportion increases. 
For OP ratios of 5\%, 20\%, 35\%, and 50\%, \(p\) stabilizes at approximately 25\%, 63\%, 84\%, and 96\%, respectively. 
That is, \uline{as the OP ratio rises from 5\% to 50\%, active object migration steadily declines, disappearing above 50\%, eliminating L2SWA(A). \textbf{(Observation 4)}}

\paragraph{Summary:}
\textit{Observations 1 and 3 show that migrating objects from HLog to HSet leads to high WA for both passive and active migration, and \(\text{L2SWA}\) can be simplified as:
\begin{equation}
  \text{\textbf{L2SWA}} = (2 - p) \cdot \text{\textbf{L2SWA(P)}} = \frac{(2 - p)\,  N'_{\text{Set}}}{2N_{\text{Log}}} 
\end{equation}
\textit{When combining with Observations 2 and 4, this formulation shows that enlarging HLog or adjusting HSet's OP ratio still fails to achieve near-minimal flash writes, even at the cost of high memory and low flash utilization. 
A closer look at the formula indicates that, since HLog must remain small, write amplification is mainly driven by the large number of sets in HSet, i.e., a large hash range. The large hash range implies a lower probability of hash collisions before migration, which results in less newly written objects per set write, hereafter referred to as \uline{a low per-set fill rate}.}}

\section{\ZNSKVCache~ Design}
\label{sec:design}
\subsection{Overview}
\label{sec:overview}
\begin{figure}[t]
    \centering
    \includegraphics[width=0.95\linewidth]{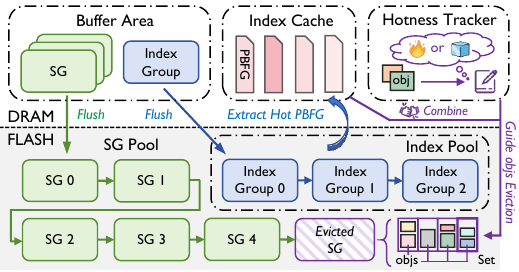}
     \vspace{-0.12in}
    \caption{The overview of \ZNSKVCache. }
    \label{fig:overview}
     \vspace{-0.2in}
\end{figure}

Based on prior analysis, we propose \ZNSKVCache, a novel flash cache architecture for tiny objects that achieves near-ideal write amplification while maintaining memory efficiency and performance. 
Figure~\ref{fig:overview} presents the overview of \ZNSKVCache.

\textit{\textbf{Minimize write amplification.}} 
\ZNSKVCache adopts two key design choices. 
First, it intentionally uses a small hash space to increase per-set fill rate, and treats the set-associative layout as a logical unit called the Set-Group (SG) composed of multiple sets.
The SG size is configurable, typically aligned to a multiple of the underlying device erase unit.
Second, it performs both flushing and eviction at the SG level, enabling flash-friendly and batched write patterns.
As shown in Figure~\ref{fig:overview}, an SG begins as a mutable in-memory structure to aggregate incoming objects. It is later flushed to become an immutable SG in the on-flash SG pool managed in a FIFO manner. 
Under this design, \ZNSKVCache's write amplification equals the reciprocal of the expected fill rate of each SG, the aggregate fill rate of its constituent sets.
\begin{equation}
  \text{\textbf{WA(Nemo)}} = \frac{1}{\mathbb{E}(\text{FR}_{\text{SG}})}
\end{equation}

Compared to FairyWREN, whose write amplification arises from high fill-rate log writes and low fill-rate set writes (Equation \ref{eq:1}), \ZNSKVCache~restructures the architecture and defines \textbf{a fourth class of flash cache design}. Specifically, \ZNSKVCache~consolidates two write sources into a single, high fill-rate sequential write, eliminating log-to-set migration and minimizing write amplification. 
It preserves the logical benefits of set-associative caching while adopting log-structured physical writes, preventing the read-modify-write overhead of traditional set-associative design.





\textit{\textbf{Index design.}} 
In \ZNSKVCache, object lookups require two steps: identifying the target SG and locating the object's intra-SG offset. 
The intra-SG offset is derived from the object's hashed key, while finding the correct SG is more challenging. 
An exact object-to-SG mapping, even with efficient indexing structures like hash table, requires over 40 bits/obj, accounting for SG ID (10+bits), next pointers
(32+bits), and tags.
In comparison, hierarchical caches (Kangaroo and FairyWREN) cost fewer than 10 bits/obj. 
To achieve the same memory efficiency, \ZNSKVCache~ proposes approximate object indexing with bloom filters and selective metadata offloading from memory to disk, without compromising performance.


Approximate object indexing in \ZNSKVCache~is feasible due to two factors: 
(1) SG-level batched writes minimize write-induced interference on reads, allowing the system to tolerate occasional false positives; 
and (2) \ZNSKVCache's set-associative layout ensures that each object can reside only in  a set within SGs, enabling tighter control of false positives than in log-structured caches. In this work, we instantiate the approximate object indexing mechanism as the Parallel Bloom Filter Group (PBFG), which aggregates Bloom Filters (BFs) from all SGs. 
The filters are gradually constructed as objects are inserted into the in-memory SGs and are configured with low false-positive rates to reduce misreads.
During lookups, the filters are queried in parallel to identify candidate SGs, followed by parallel reads of the corresponding sets at the hashed key offset. 
As a result, PBFG achieves significant memory savings; for example, with a 1\% false-positive rate, the indexing structure consumes only 9.6 bits per object.

For the selective metadata offloading, \ZNSKVCache~only keeps frequently updated metadata in memory, while storing rarely updated metadata on flash and caching its hot entries in memory. 
In this work, the metadata contains the bloom filters used to form PBFG, and the access counter used to track object hotness. 
The former is stable, as it is updated only before the in-memory SG is flushed to flash, and will not change until the SG is evicted. 
In contrast, the latter is frequently updated, with each object access triggering an update.
To illustrate this process, Figure \ref{fig:overview} shows that the bloom filter for incoming objects is buffered and updated within the in-memory index group, then offloaded to the on-flash index pool. Meanwhile, in memory, all object's access counters (hotness tracker) are maintained, and the frequently accessed objects' PBFGs are cached in the index cache.

\textit{\textbf{Object hotness tracking.} }
Extending these methods, we make a further observation: the index cache implicitly encodes object hotness, as it tends to retain the PBFGs of recently accessed objects. Based on this insight, \ZNSKVCache~ adopts a hybrid tracking mechanism that combines low-bit access counters with index cache state to infer object hotness, significantly reducing memory overhead. 

\textit{\textbf{Operations.}}
Three cache operations are available in \ZNSKVCache.
\ding{182} \textbf{Insert}: upon receiving an object, \ZNSKVCache~hashes its key to locate the corresponding set within the in-memory SG, buffers the object, and updates the BF in the in-memory index group. 
When the buffers are full, they are flushed to the on-flash pool.
\ding{183} \textbf{Lookup}: \ZNSKVCache~performs a hierarchical search. 
It first calculates the intra-SG offset from the hashed key to check whether the object is in the in-memory SG. 
If not found, it queries the index cache for the matching PBFG. 
If the PBFG is present in memory, \ZNSKVCache~derives the candidate SGs and accesses them in parallel. 
If the PBFG is absent, \ZNSKVCache~retrieves it from the index pool from flash and loads it into the in-memory index cache for further access.
\ding{184} \textbf{Eviction}: when the on-flash SG pool reaches its maximum capacity, it evicts the earliest SG. During eviction, \ZNSKVCache~writes back hot objects to the in-memory SG.

We identify three challenges (C1{\textasciitilde}C3) to implement \ZNSKVCache. 
Following the analysis of each, we present three solutions.

\begin{figure}[t]
    \centering
    \begin{subfigure}[b]{0.222\textwidth}
        \includegraphics[width=0.9\linewidth]{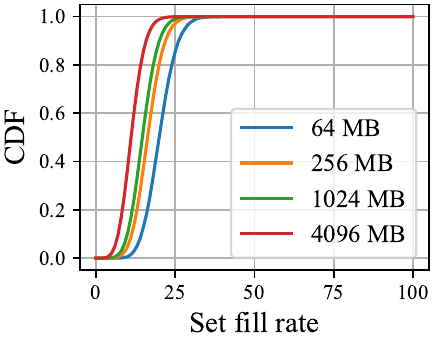}
        \vspace{-0.08in}
        \caption{Synthetic 4 KB insert.}
        \label{fig:design_cdf4}
    \end{subfigure}
    \hfill
    \begin{subfigure}[b]{0.222\textwidth}
        \includegraphics[width=0.9\linewidth]{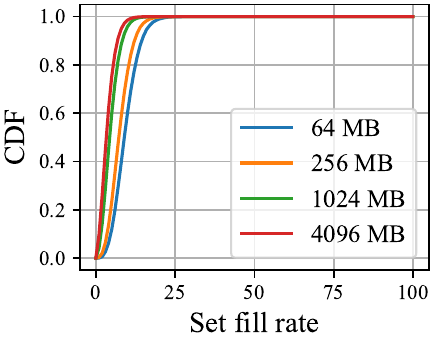}
        \vspace{-0.08in}
        \caption{Real-world 4 KB insert.}
        \label{fig:design_cdf4_r}
    \end{subfigure}
    \hfill
    \begin{subfigure}[b]{0.222\textwidth}
        \includegraphics[width=0.9\linewidth]{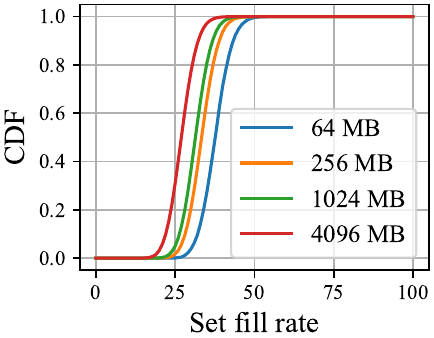}
        \vspace{-0.08in}
        \caption{Synthetic 8 KB insert.}
        \label{fig:design_cdf8}
    \end{subfigure}
    \hfill
    \begin{subfigure}[b]{0.222\textwidth}
        \includegraphics[width=0.9\linewidth]{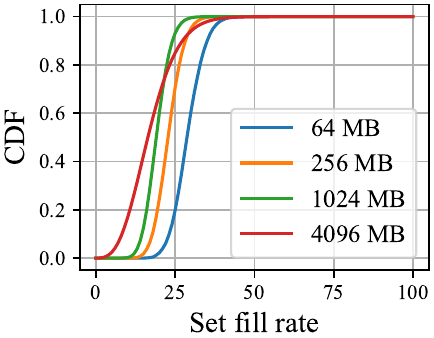}
        \vspace{-0.08in}
        \caption{Real-world 8 KB insert.}
        \label{fig:design_cdf8_r}
    \end{subfigure}
    \vspace{-0.17in}
    \caption{Short-term hashed key distribution skew.}
    \label{fig:design_cdf}
    \vspace{-0.3in}
\end{figure}

\textit{\textbf{C1: How to prepare a ``perfect'' SG before flush?}}
A ``perfect'' SG indicates the internal sets are populated with high fill rates; if such SGs are flushed, {\ZNSKVCache} can deliver low ALWA.
In naïve {\ZNSKVCache}, the timing for flushing a SG from memory to flash is when any set within that SG is full.
In practice, a key design challenge arises during SG population: regardless of the specific hash function used, the hashed key distribution can exhibit short-term skew.
To quantify this, we tested SGs of varying sizes (64 MB to 4096 MB) and set sizes (4 KB and 8 KB) using both real-world Twitter traces and synthetic workloads (with data sizes following a normal distribution, mean = 250 B, std = 200 B). 
As shown in Figure \ref{fig:design_cdf}, when a set is first filled, the remaining sets within the SG tend to have low fill rates. 
In the 4 KB configuration, the fill rate of remaining sets is typically below 25\%, regardless of the workload, and even with an 8 KB set size, it rarely exceeds 40\%. 
This reveals that, when an SG is populated from empty, once any set is full, the remaining sets remain sparsely populated. 
Directly flushing these underfilled SGs will result in significant write amplification and wasted cache capacity. 
Although increasing the set size can reduce skew, it comes at the cost of higher read amplification.
Consequently, for minimal write amplification, preparing a highly filled SG, i.e., a ``perfect'' SG, before each flush is challenging.

\textit{\textbf{C2: How to enable efficient offloading?}}
Without careful design, the BFs forming the PBFGs cannot be efficiently offloaded to flash.
In a naïve implementation, each SG maintains a BF, and all objects inserted into the SG are recorded by the SG-level BF. 
Thus, \ZNSKVCache~ has only a single PBFG composed of SG-level BFs of all SGs.
Thus, deriving the candidate SGs for any object requires the entire PBFG to be in memory, making index offloading to disk impractical.
To enable the offloading, there are three conditions must be satisfied:
(1) There must be multiple PBFGs, each of which can be independently queried for locating one object's candidate SGs;
(2) These PBFGs must exhibit sufficient hotness skew, allowing on-demand caching of indexing metadata;
(3) The I/O cost of retrieving a PBFG from flash must remain low.

\textit{\textbf{C3: How to exploit coarse-grained index hotness for efficient eviction?}}
Designing hybrid hotness tracking mechanism faces two key issues. 
First, since a PBFG covers a large number of objects, its hotness represents an aggregated, coarse-grained access pattern of these objects. 
It is therefore difficult to reliably correlate this PBFG-level hotness with the fine-grained hotness of individual objects for accurate hotness identification.
Second, as the SSD capacity increases, objects remain in flash for longer durations, and their access patterns change over time. 
This makes it hard to distinguish whether an object's hotness stems from an initial burst (now cooled) or long-term popularity, thus posing a challenge for using a low-bit access counter.

\begin{figure}[t]
    \centering
    \includegraphics[width=0.99\linewidth]{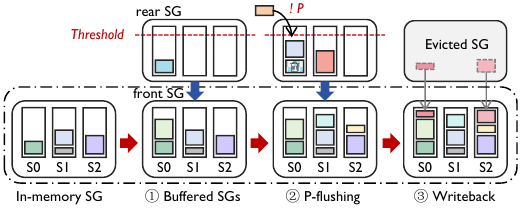}
    \vspace{-0.1in}
    \caption{Three techniques for increasing fill rate per SG.}
    \label{fig:design_fill_rate}
     \vspace{-0.25in}
\end{figure}

\subsection{Preparing a ``Perfect'' Set-Group (C1)}
\label{sec:perfect_SG}
To solve challenge 1, \ZNSKVCache~ strategically delays the time of flushing in-memory SGs to flash at low cost, thereby alleviating the impact of short-term hash skew.
As shown in Figure \ref{fig:design_fill_rate}, this is achieved through three key techniques:

\textit{Buffered in-memory SGs.}
\ZNSKVCache~ adopts a buffered in-memory SGs structure organized as a circle queue, as shown in the \textcircled{1} of Figure \ref{fig:design_fill_rate}. 
When an object is received, \ZNSKVCache~ inserts it into the set of the available SG closest to the queue's front.
If multiple in-memory SGs exist, this approach can delay the front SG flushing, allowing those underutilized sets to continue absorbing more objects. 
The front SG will be flushed once the rear SG is nearly full.
This indicates that allocatable space is almost exhausted and a flush is necessary to reclaim room for new insertions. 
During SG flushing, new objects are inserted into available SG, and a new empty SG will be added to the queue as a new rear SG after flushing. 
This design decouples the SG flushing and cache insertion, ensuring continuous object insertion without blocking.

\textit{Probabilistic flushing mechanism.} 
\ZNSKVCache~ employs a probability parameter $p$ to further delay the SG flushing, as shown in the \textcircled{2} of Figure \ref{fig:design_fill_rate}.
A random value $p$ will be generated; if $p$ falls below a predefined threshold $p_{th}$, the front SG will be flushed immediately. 
Otherwise, the flush is held, and room will be made for new objects by evicting objects from the sets corresponding to their hashed key. 
For example, with $p_{th}=0.001$, only about an immediate flush in 1,000 flush attempts.
In practice, each immediate flush incurs the cost of evicting roughly 1,000 objects, while the benefit is the insertion of up to millions of new objects into the front SG.
This favorable trade-off allows \ZNSKVCache~ to achieve a substantially higher overall SG fill rate at minimal eviction cost.

\textit{Hotness-aware writeback during SG eviction.} 
Once flash cache is nearly full, each flushing of an in-memory SG triggers an eviction of the earliest SG from the on-flash SG pool. 
\ZNSKVCache~ re-inserts hot objects from the on-flash evicted SG into the to-be-flushed in-memory SG to further fill the sets, thereby increasing SG fill rate and ensuring that hot objects remain in cache (Figure \ref{fig:design_fill_rate} \textcircled{3}).
During the re-insertion process, the to-be-flushed SG no longer accepts new insertions but provides read access.

Based on these techniques, \ZNSKVCache~ gradually constructs a ``perfect'' SG, characterized by a high fill rate of 89.34\%, which allows \ZNSKVCache~to achieve low WA and high cache utilization.

\begin{figure}
    \centering
    \includegraphics[width=0.95\linewidth]{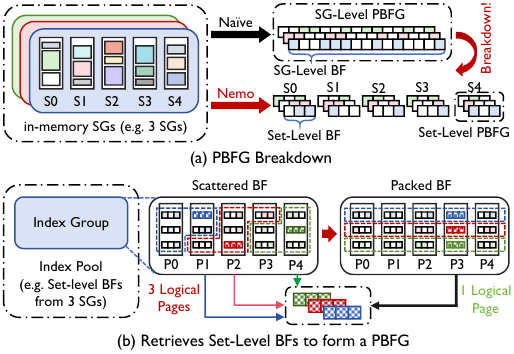}
     \vspace{-0.12in}
    \caption{PBFG breakdown and retrieving a PBFG from flash. ``S0$\sim$S4'' refers to the 5 sets in an SG and ``P0$\sim$P4'' represents the 5 pages in an index group on the flash.}
    \label{fig:design_scatter}
     \vspace{-0.17in}
\end{figure}

\subsection{Lightweight Index Structure (C2)}
\label{sec:lightweight_index}
To solve challenge 2, \ZNSKVCache~ builds a lightweight indexing mechanism, which attempts to persist all indexes and only caches a small portion in memory.
Here we apply three optimizations.
First, we break down the naïve SG-level PBFGs into fine-grained Set-level PBFGs.
This breakdown is driven by two facts: (1) Sets within an SG are independent and can be queried directly via object key.
(2) The space efficiency of a bloom filter depends only on its target false-positive rate, not on the number of its member objects. 
Thus, a single SG-level BF can be broken down into multiple Set-level BFs, each dedicated to a single set, without sacrificing space efficiency. 
As shown in Figure \ref{fig:design_scatter}(a), all Set-level BFs corresponding to the same intra-SG offset across different SGs are grouped into a Set-level PBFG. Thus, lookups and insertions in \ZNSKVCache~ are performed as follows.
For Lookups, \ZNSKVCache~ queries all Set-level BFs in a PBFG in parallel, quickly finding out all candidate SGs that may contain the object. 
For insertions, \ZNSKVCache~ places objects into its target set and updates the corresponding Set-level BF accordingly.

Second, we exploit PBFG hotness to support on-demand indexing caching.
Here, the hotness of a PBFG refers to the aggregated access frequency of its member objects. 
Since object access pattern typically follows a Zipfian distribution \cite{yang2021large}, the access frequency of sets remains highly skewed.
Consequently, retaining only the most popular PBFGs in memory is enough to serve most object accesses.

Finally, to minimize the I/O overhead of loading Set-level PBFGs from flash, we optimize Set-level PBFGs' physical placement on flash.
The key principle is to align the Set-level PBFG's granularity with the flash page.
In a naïve design, the Set-level BFs of a specific SG are built sequentially and flushed alongside its SG. 
While simple, this approach scatters Set-level BFs of a single Set-level PBFG across multiple non-contiguous pages. 
As shown in Figure~\ref{fig:design_scatter}(b), the Set-level BFs (e.g., dot-filled boxes) of a Set-level PBFG are distributed across 3 pages. 
Thus, retrieving this Set-level PBFG needs to read 3 pages (P1, P2, and P4), incurring high read amplification.
To address this, \ZNSKVCache~ slightly enlarges the in-memory index group so that more Set-level BFs from different SGs can be constructed and form multiple Set-level PBFGs, each sized to fit within a single flash page.
For example, in the right side of Figure~\ref{fig:design_scatter}(b), the packed BF layout indicates that P0$\sim$P4 contain five Set-level PBFGs. 
In this case, retrieving a Set-level PBFG across three SGs only requires reading one page (P3). 
In practice, a 4 KB flash page can accommodate tens or even hundreds of Set-level BFs, substantially reducing the I/O cost of Set-level PBFG retrieval.

\subsection{Hybrid Hotness Tracking Mechanism (C3)}
\label{sec:hybrid_hotness}
\begin{figure}
    \centering
    \includegraphics[width=0.95\linewidth]{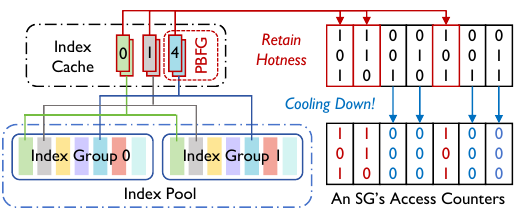}
    \vspace{-0.13in}
    \caption{Cooling object hotness within an SG.}
    \label{fig:design_hotness}
     \vspace{-0.23in}
\end{figure}

To solve challenge 3, \ZNSKVCache~ leverages the hotness features of PBFG and the access counter.
In detail, \ZNSKVCache~ exploits the recency feature of the cached PBFGs at set level and the frequency feature of the access counter for objects within those sets. 
An object is considered ``hot'' if it exhibits high access frequency and its associated set is recently active (indicated by a cached PBFG).
In our implementation, access counters are single-bit entries, realized as a bitmap.
As shown in Figure \ref{fig:design_hotness}, when PBFG 0, 1, and 4 are cached (indicating their sets as recently active), the corresponding bits in the access counter are checked to determine if the associated objects are truly ``hot''.
Once identified, when a victim SG is picked for eviction after the cache is full, hot objects will be retained by re-insertion, and cold objects will be evicted.

To address evolving object hotness, \ZNSKVCache~ employs a periodic cooling mechanism. 
Cooling is periodically triggered, which resets all corresponding bits in the access counter. 
As shown in Figure \ref{fig:design_hotness}, bits for sets with cached PBFGs (e.g., 0, 1, 4) are retained, while others are cleared. 
This process repeats periodically throughout an object’s lifetime, ensuring that only recency-backed hotness is sustained.
As hotness tracking is primarily used to guide eviction decisions by distinguishing hot and cold objects, \ZNSKVCache~ records hotness information only during an object's later-life stage, significantly reducing the memory overhead of the tracking mechanism.

\section{Evaluation}
\label{sec:eval}
Our evaluation contains four aspects: 
(1) \textbf{Main Metrics (\S \ref{res:wa_miss}):} write amplification, read latency and miss ratio.
(2) \textbf{Design Breakdown (\S \ref{res:breakdown}):} the contribution of each design component.
(3) \textbf{Sensitivity Analysis (\S \ref{res:sen}):} performance under varying configuration parameters.
(4) \textbf{Overhead (\S \ref{res:overhead}):} memory usage, PBFG computational overhead and read amplification.

\subsection{Methodology} 
\label{sec:methodology}

\textbf{Hardware configuration.} 
Experiments were conducted on a server equipped with a 24-core Intel CPU, 128 GB of memory, running Ubuntu 22.04 with Linux kernel version 5.15, and using a Western Digital Ultrastar DC~ZN540 ZNS SSD \cite{ZN540}, where each zone has a capacity of 1,077 MB.

\begin{table}[t]
\centering
\small
\caption{Nemo configuration. The flushing threshold is count-based, not probabilistic.}
 \vspace{-0.1in}
\begin{tabular}{cc}
\toprule
\textbf{Parameter} & \textbf{Value} \\
\midrule
Set size & 4 KB \\
Sets per SG & 275{,}712 \\
PBFG false positive rate & 0.1\% \\
\# SGs : \#  index groups & 50:1 \\
\midrule
\# in-memory SGs & 2 \\
Flushing threshold ($p_{th}$) & 4,096 \\
Cached PBFG ratio & 50\% \\
Hotness tracking start position & Last 30\% of cache \\
SG cooling period & Every 10\% cache written \\
\bottomrule
\end{tabular}
\label{tab:nemo-config}
\vspace{-0.1in}
\end{table}

\textbf{Implementation.}
\label{sec:imple}
\ZNSKVCache~ is built as a cache module in CacheLib \cite{berg2020cachelib}. Background tasks (e.g., SG flushing and object writeback) run on dedicated asynchronous threads to avoid blocking foreground requests. To improve concurrency, we use fine-grained locks: each in-memory set has its own read-write lock, allowing parallel inserts; writeback only locks the affected SG, so reads can still proceed; each PBFG in the index cache has a read-write lock to reduce access latency. The index cache is FIFO-style, which reduces lock contention under high access pressure compared to LRU \cite{ qiu2024can, yang2023fifo}.

To enable \ZNSKVCache~on the ZN540, we use the \texttt{libzbd} library \cite{libzbd} to interact with it. Before running the evaluation, we tuned several parameters to fit the hardware. The SG size was set to match the zone capacity, with 275,712 sets per SG, and each set fixed at 4 KB (= a flash page). This large SG flushing helps make better use of ZNS parallel I/O bandwidth. The bloom filters in PBFG are configured with a 0.1\% false positive rate, corresponding to 14.4 bits/obj. When targeting 40 objects per set, each filter requires only 576 bits (72 bytes), allowing 50 filters to fit in a single flash page. As a result, each index group stores bloom filters for 50 SGs. Additional configurable parameters are listed in Table \ref{tab:nemo-config}.

\textbf{Baselines.} 
We compare \ZNSKVCache~ against four baselines: log-structured cache (\textbf{Log}), set-associative cache (\textbf{Set}), FairyWREN (\textbf{FW}), and Kangaroo (\textbf{KG}). All systems are implemented as CacheLib cache engines and evaluated within the same CacheLib framework. The Log and Set baselines are natively supported by CacheLib. For FairyWREN, we use a version with known bugs fixed. 
For Kangaroo, as it lacked ZNS interface support, we added it to the released code \cite{KG-sourcecode}. 
The experimental parameters are summarized in Table \ref{tab:parameter}, and all baselines are evaluated on WD ZN540.

\begin{table}[t]
\setlength{\tabcolsep}{2mm}{
\centering
\small
\caption{Experimental parameters of cache engines.}
\label{tab:parameter}
 \vspace{-0.1in}
\begin{tabular}{c|ccccc}
\toprule
\textbf{Parameter} & \textbf{\ZNSKVCache} & \textbf{Log} & \textbf{Set} & \textbf{FW} & \textbf{KG}\\
\midrule
Flash size (GB) & 360 & 360 & 360 & 360 & 360\\
OP space (GB) & 2 & 2 & 200 & 18 & 18 \\
Log of cache size  & 0\% & 100\% & 0\% & 5\% & 5\% \\
Set of cache size  & 100\% & 0\% & 100\% & 95\% & 95\% \\
\bottomrule
\end{tabular}}
 \vspace{-0.1in}
\end{table}

\begin{table}[h]
\setlength{\tabcolsep}{1.mm}{
\centering
\small
\caption{Characteristics of Twitter traces.}
\label{tab:trace}
 \vspace{-0.1in}
\begin{tabular}{ccccc}
\toprule
\textbf{Trace} & \textbf{K-Size (B)} & \textbf{V-Size (B)} & \textbf{WSS (MB)} & \textbf{Zipf $\alpha$} \\
\midrule
cluster\_14 & 96 & 414 & 18333 & 1.2959 \\
cluster\_29 & 36 & 799 & 40520 & 1.2323 \\
cluster\_34 & 33 & 322 & 11552 & 1.1401 \\
cluster\_52 & 20 & 273 & 14057 & 1.2117 \\
\bottomrule
\end{tabular}}
\vspace{-0.1in}
\end{table}

\begin{figure*}[t]
  \centering
  \begin{minipage}[b]{0.385\textwidth}
    \begin{subfigure}[b]{0.48\textwidth}
        \includegraphics[width=\linewidth]{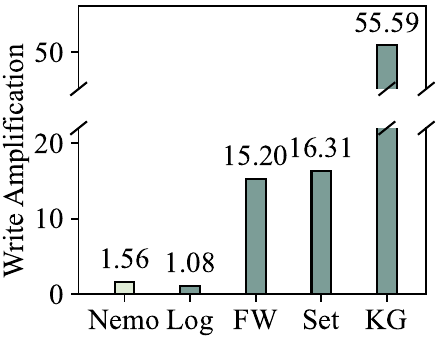}
        \caption{WA of five systems}
        \label{fig:res_wa_four_baseline}
    \end{subfigure}
    \hfill
    \begin{subfigure}[b]{0.49\textwidth}
        \includegraphics[width=\linewidth]{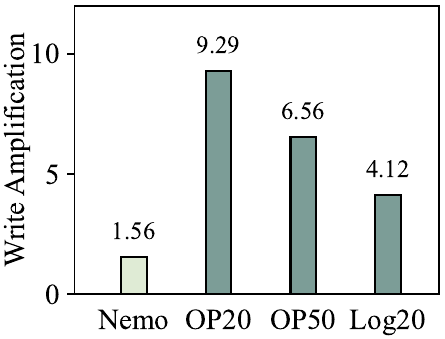}
        \caption{\ZNSKVCache~ vs. FWs}
        \label{fig:res_wa_fw_config}
    \end{subfigure}
     \vspace{-0.1in}
    \caption{WA of different cache systems.}
    \label{fig:res_wa}
  \end{minipage}
  \hfill
  \begin{minipage}[b]{0.285\textwidth}
    \includegraphics[width=\linewidth]{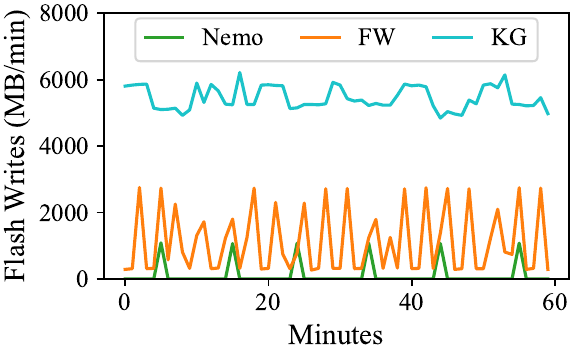}
     \vspace{-0.23in}
    \caption{Flash write patterns.}
    \label{fig:res_flash_write}
  \end{minipage}
  \hfill
  \begin{minipage}[b]{0.32\textwidth}
    \includegraphics[width=\linewidth]{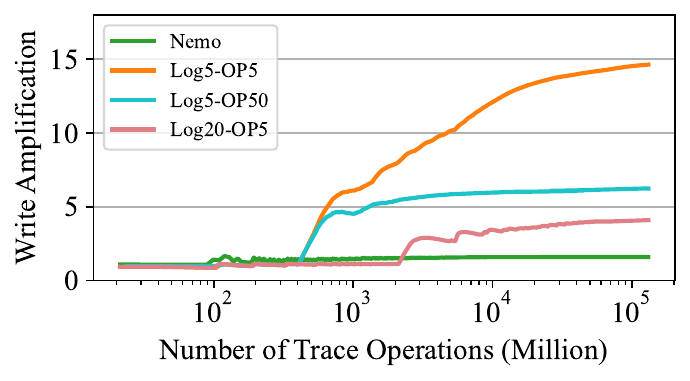}
     \vspace{-0.25in}
    \caption{WA trends with operations.}
    \label{fig:res_wa_all_trace}
  \end{minipage}
  \vspace{-0.25in}
\end{figure*}

\begin{figure*}[t]
     \vspace{-0.05in}
    \centering
    \begin{subfigure}[b]{0.33\textwidth}
        \includegraphics[width=\linewidth]{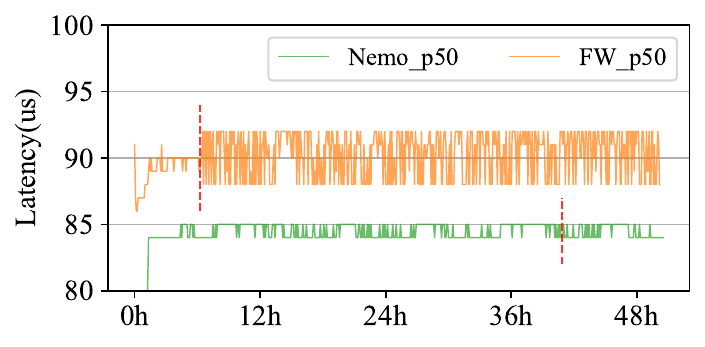}
         \vspace{-0.25in}
        \caption{p50 latency comparision.}
        \label{fig:res_p50}
    \end{subfigure}
    \hfill
        \begin{subfigure}[b]{0.33\textwidth}
        \includegraphics[width=\linewidth]{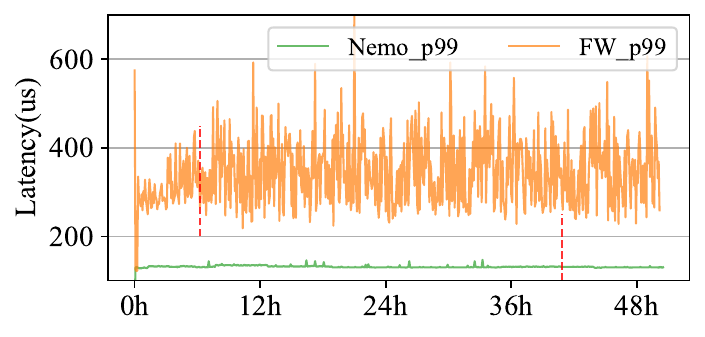}
         \vspace{-0.25in}
        \caption{p99 latency comparision.}
        \label{fig:res_p99}
    \end{subfigure}
    \hfill
    \begin{subfigure}[b]{0.33\textwidth}
        \includegraphics[width=\linewidth]{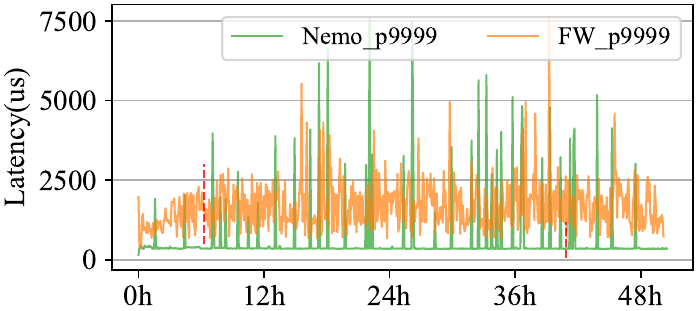}
         \vspace{-0.2in}
        \caption{p9999 latency comparison.}
        \label{fig:res_p9999}
    \end{subfigure}
    \vspace{-0.28in}
    \caption{Read latency of Nemo and FW.}
    \label{fig:res_latency}
     \vspace{-0.13in}
\end{figure*}

\textbf{Benchmarks.} 
Several traces from Twitter real platforms\footnote{Meta traces \cite{meta_trace} are dominated by large objects, and persisting large objects in the flash cache does not constitute a performance bottleneck.} are evaluated in this work \cite{yang2021large}.
Four representative traces (clusters 14, 29, 34, and 52) are selected based on three criteria. 
First, the traces exhibit large working set sizes (WSS $>$ 8~GB) to trigger evictions. 
Second, their access patterns follow a Zipfian distribution ($\alpha \approx 1$), capturing the skew typical of real-world KV cache workloads. For reference, $\alpha = 0$ corresponds to uniform access, $\alpha = 1$ represents the classic 80/20 Pareto distribution, and higher values indicate increasingly skewed workloads \cite{newmanPowerParetoZipf, atikoglu2012workload, zipf2016human}.
Third, the traces' average object size is small.
The key characteristics of these traces are summarized in Table~\ref{tab:trace}.
To apply realistic cache pressure, we scale and merge the traces.
Specifically, we increase the working set size by running each trace concurrently across four disjoint key spaces, and proportionally interleave requests from the four traces to avoid periods dominated by a single workload’s characteristics.
Following the average object size in FairyWREN and Kangaroo, we downscale object sizes by 2$\times$ and 3$\times$ for clusters 14 and 29, respectively, while leaving clusters 34 and 52 unchanged.
This results in an average object size of 246~B (265~B/271 B in FairyWREN/Kangaroo).

\subsection{The Main Metrics of Flash Cache}

\label{res:wa_miss}
This section presents the results of several key flash cache metrics: WA, read latency, and miss ratio.  Before explaining WA, we first clarify an important point: since objects written back during the write-back operation can not be counted as logical writes, \ZNSKVCache's WA does not directly equal the inverse of the SG fill ratio when write-back is enabled. Instead, WA is calculated as the size of the SG divided by the total size of objects newly written by the first two techniques (including objects evicted early during ``probabilistic flushing'') in \S \ref{sec:perfect_SG}.

Figure \ref{fig:res_wa_four_baseline} reports the steady-state WA of the five cache systems. 
Log and Set represent two extremes. Log achieves the lowest WA of 1.08 at the cost of the highest memory overhead (>100 bits/obj). 
Set, in contrast, incurs the highest WA of 16.31 with the lowest memory cost (4 bits/obj), and reduces the usable flash capacity by half (OP = 50\%).
FW represents the SOTA trade-off in flash cache, enabling a WA of 15.2 and reducing the set OP ratio from 50\% to 5\%, with a memory overhead of 9.9 bits/obj (The memory overhead analysis is detailed in \S \ref{res:overhead}).
KG also adopts a hierarchical design, but its WA reaches 55.59. 
Compared to FW, KG lacks cold–hot partitioning, causing its set-associative back tier to have a hash range twice as large as that of FW. KG's log-to-set object migration leads to WA exceeding 15.
Additionally, its garbage collection and log-to-set object migration are independent, causing the overall WA to increase multiplicatively. When KG is stable, the number of valid sets in each erased unit is about 50\% to 80\%, meaning that the WA for each erased unit is between 2$\times$ and 5$\times$. Under the default configuration, \ZNSKVCache~ achieves a WA of 1.56. \ZNSKVCache~ reduces flash writes by 9$\times$ compared to FW with a memory overhead of 8.3 bits/obj and requires less than 1\% OP ratio.
Moreover, we measured flash writes per minute for Nemo, FW and KG at steady state (Figure \ref{fig:res_flash_write}). The results show that Nemo only incurs occasional small writes, while FW and KG experience continuous writes, with KG's flash writes per minute significantly higher than FW's. Additionally, Nemo performs batched writes, whereas FW and KG's writes are almost entirely set-level requests.

\begin{figure*}[t]
  \centering
  \begin{minipage}[b]{0.27\textwidth}
    \includegraphics[width=\linewidth]{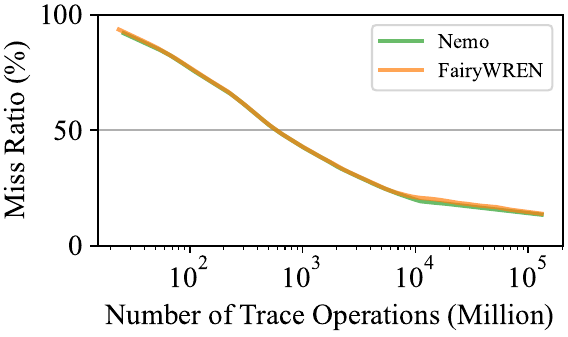}
     \vspace{-0.25in}
    \caption{Miss ratio trend.}
    \label{fig:res_miss}
  \end{minipage}
  \hfill
  \begin{minipage}[b]{0.29\textwidth}
    \includegraphics[width=\linewidth]{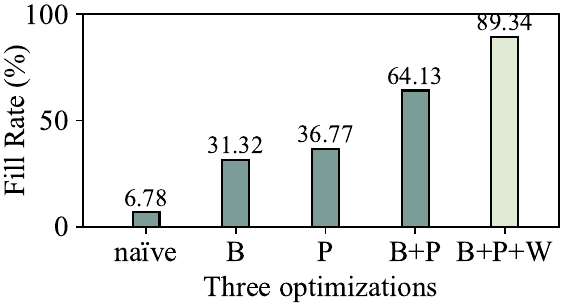}
     \vspace{-0.25in}
    \caption{``Perfect'' SG breakdown.}
    \label{fig:res_fill_rate}
  \end{minipage}
  \hfill
  \begin{minipage}[b]{0.42\textwidth}
    \includegraphics[width=\linewidth]{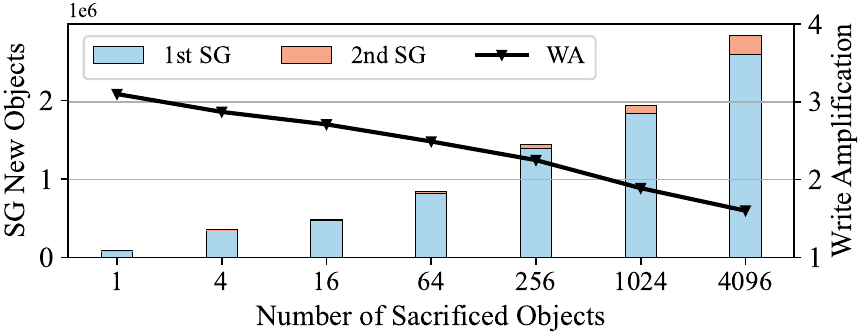}
     \vspace{-0.25in}
    \caption{Benefits of P-flushing on buffered SGs.}
    \label{fig:res_randomp_objNum}
  \end{minipage}
  \vspace{-0.1in}
\end{figure*}

We further compare \ZNSKVCache~ and FW under various configurations. 
There are two methods to reduce the WA of FW: expanding the HLog size or increasing the OP ratio of HSet. Accordingly, we enlarge the HLog size from 5\% to 20\% of the flash capacity (incurring a 73\% increase in memory usage) and increase the OP ratio from 5\% to 50\% (nearly halving the effective cache capacity), and measure the WA after the systems are stable. As shown in Figure~\ref{fig:res_wa_fw_config}, FW with 20\% HLog and 5\% OP ratio (\uline{Log20}-OP5) and FW with 5\% HLog and 50\% OP ratio (Log5-\uline{OP50}) still yield much higher WA than \ZNSKVCache, at 4.12 and 6.56, respectively.

We also trace the variation of these systems' WA with the number of trace operations.
As shown in Figure \ref{fig:res_wa_all_trace}, \ZNSKVCache~ stabilizes WA at around 1.56. FW initially achieves a lower WA of 1.1 when only HLog is active. However, as the number of operations increases, FW exhibits a sharp rise in WA, since the HLog becomes exhausted and passive object migration is triggered. For Log5-OP5 and Log20-OP5, the timing of the first sudden increase of WA differs, as the 5\% HLog is consumed more quickly than the 20\% HLog. 
Both configurations show another distinct turning point, corresponding to active migration. 
In contrast, when comparing Log5-OP5 with Log5-OP50, WA grows more slowly in Log5-OP50 after passive migration begins, since its HSet hash range is narrower. At later stages, Log5-OP50 triggers GC earlier due to its smaller cache capacity. 
A key difference is that Log5-OP50 does not exhibit a second turning point, because active object migration rarely occurs in this configuration.

To assess system stability, we compare the read latency of Nemo and FW before and after the flash space is fully utilized (as indicated by the red dashed line in the figure). As shown in Figure \ref{fig:res_latency}, Nemo consistently outperforms FW across $p50$, $p99$, and $p9999$ latency, demonstrating better stability. At the $p50$ latency, both systems remain stable, with Nemo offering a 5 µs advantage. 
For tail latency, Nemo's $p99$ and $p9999$ latencies remain stable at around 131 µs and 523 µs, respectively, while FW experiences significant fluctuations, with $p99$ varying drastically around 350 µs and $p9999$ fluctuating around 1488 µs.
The higher and more erratic latency in FW is due to frequent 4 KB write operations that disrupt subsequent reads.  In contrast, Nemo's batching write pattern minimizes interference, ensuring consistently lower, more stable, and predictable read performance, crucial for maintaining high-quality data services.

Finally, Nemo and FW exhibit similar miss ratios (Figure \ref{fig:res_miss}), as Nemo’s hotness-aware writeback mechanism keeps hot objects in the cache, and the working set of hot data is smaller than the cache space for both systems.

\subsection{What Makes \ZNSKVCache~ Better}
\label{res:breakdown}
\ZNSKVCache~ employs three techniques to improve the SG fill rate, and their individual contributions are evaluated separately.
The impact of \textit{buffered in-memory SGs} (B), \textit{probabilistic flushing} (P), and \textit{hotness-aware writeback} (W) technique is examined through controlled experiments, as summarized in Figure~\ref{fig:res_fill_rate}.
The naïve version of \ZNSKVCache~(without any optimization) achieves only a 6.78\% fill rate, primarily because the skewed hash function causes the front SG to saturate quickly without sufficient aggregation.
With \textit{buffered in-memory SGs}, the fill rate rises to 31.32\%.
The \textit{probabilistic flushing}, configured with $p_{th}$ = 4096, improves the fill rate to 36.77\%.
These two mechanisms both delay SG flushing, thus providing more time for aggregation.
When combined (B+P), the two techniques achieve a 64.13\% fill rate, with Nemo's ALWA approximately equal to the reciprocal of the fill rate.
Finally, with \textit{hotness-aware writeback}, the full version (B+P+W) achieves the highest fill rate of 89.34\%, representing an additional 25.21\% improvement over (B+P).

\subsection{Sensitivity Analysis}
\label{res:sen}
We conduct an experimental analysis on the sensitivity of \ZNSKVCache~ to configurable parameters. Initially, we evaluate the optimization effects of the \textit{probabilistic flushing} with different $p_{th}$, focusing on fill rate gains and WA (Figure \ref{fig:res_randomp_objNum}). 
The results indicate that as the value of the $p_{th}$ increases, the number of new objects also rises, while \ZNSKVCache's WA decreases. However, the mechanism's profit, defined as the ratio of new objects to evicted objects, exhibits diminishing returns. For instance, when the $p_{th}$ value increased from 64 to 1024, the number of new objects only doubled.

\begin{figure}[t]
    \centering
    \begin{subfigure}[b]{0.23\textwidth}
        \includegraphics[width=\linewidth]{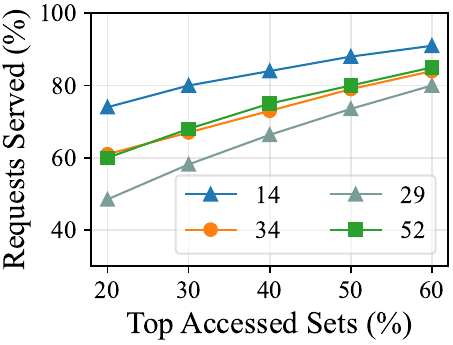}
         \vspace{-0.2in}
        \caption{Set access distribution.}
        \label{fig:res_set_hit}
    \end{subfigure}
    \hfill
    \begin{subfigure}[b]{0.22\textwidth}
        \includegraphics[width=\linewidth]{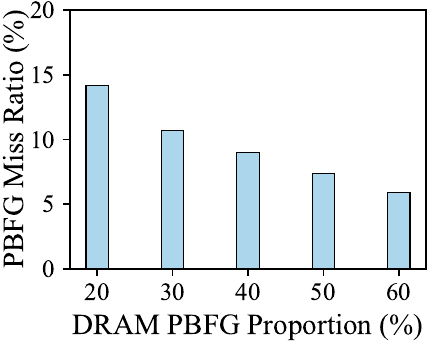}
         \vspace{-0.2in}
        \caption{PBFG misses in \ZNSKVCache.}
        \label{fig:res_pbfg_miss}
    \end{subfigure}
    \vspace{-0.12in}
    \caption{Evaluation of intra-SG offset access frequency and PBFG retrieval from index pool.}
     \vspace{-0.15in}
\end{figure}

Next, we evaluate the access frequency of intra-SG offsets and the proportion of PBFGs retrieved from the on-flash index pool. A prior experiment tested the cumulative access counts of hashed keys to their corresponding intra-SG offsets across four clusters (Figure \ref{fig:res_set_hit}). The results show that while object access frequency skew is reduced after hashing, the set access distribution still exhibits a strong skew, with $\approx$70\% of hot accesses concentrated in the top 30\% of hot sets. Further testing in \ZNSKVCache~ (Figure \ref{fig:res_pbfg_miss}) reveals that, across all in-memory PBFG ratios, fewer than 15\% of requests require PBFGs from the index pool. At a 50\% in-memory PBFG ratio, less than 8\% of requests access PBFGs from flash memory.

\subsection{Overhead}
\label{res:overhead}

\textbf{Memory usage.} As shown in Table \ref{tab:mem_overhead}, Nemo's memory overhead arises from three primary components. 
First, bloom filters are used to determine candidate SGs for each object. 
Configured with a 0.1\% false-positive rate, these filters incur 14.4 bits/obj. By retaining only 50\% of them in memory, Nemo reduces this to 7.2 bits/obj. 
Second, the hotness tracking mechanism maintains 1-bit flag for only a subset of objects (e.g., last 30\%), lowering the average cost to 0.3 bits/obj. 
Third, the memory is used to buffer SGs and the index group. Nemo's SG buffer reuses the existing memory cache, adding no overhead. The index group buffer consumes 1077 MB, equivalent to the size of a single SG. On a 2 TB device with 200 B objects, this adds 0.8 bits/obj. In total, Nemo incurs a memory overhead of 8.3 bits/obj (7.2 + 0.3 + 0.8).

\textbf{PBFG computational overhead.} Nemo leverages PBFGs to identify candidate SGs.
During lookup, each hash function is computed once and the results are shared across all filters in the PBFG, keeping the overhead low. Our measurements on ``GoogleTest'' \cite{googletest} show that computing the candidate SGs through PBFGs (including 1,000 Set-level BFs) takes about 1 µs, which is negligible compared to the tens to hundreds of microseconds required for a full lookup.

\textbf{Read amplification.} Compared with FairyWREN, Nemo achieves low lookup latency despite incurring more flash reads. We count all read activities in both systems, including reads for objects, index accesses, and reads triggered by write-back. The result shows that Nemo's read amplification is more than three times that of FW. Nevertheless, Nemo's lookup latency remains low and stable. This is mainly due to two reasons. First, most extra reads in Nemo can be parallelized, including object access from candidate SGs and retrieving PBFGs from the index pool (the worst case in lookup). Second, Nemo's write pattern introduces less interference with read requests (unlike FairyWREN's continuous random small writes), thereby preserving read performance.

\begin{table}[t]
\centering
\caption{Comparison of metadata overhead (bits per object). HLog utilizes only ~5\% of the flash space and further compresses metadata field widths.}
\vspace{-0.1in}
\small
\begin{tabular}{lccc}
\toprule
\textbf{Component} & \textbf{FairyWREN} & \textbf{Naïve Nemo} & \textbf{Nemo} \\
\midrule
Log total      & \textit{$48$ bits/obj} & -- & -- \\
\midrule 
Set index      & $\approx 3.1$ b  & $14.4$ b   & $7.2$ b (50\%) \\
Set (other)    & $3$ b & -- & -- \\
Evict          & $1$ b & $16$ b & $0.3$ b (30\%) \\
Set total      & \textit{$7.1$ bits/obj} & \textit{$30.4$ bits/obj} & \textit{$7.5$ bits/obj} \\
\midrule 
Additional       & $\approx 0.8$ b & -- & $\approx 0.8$ b \\
Log size       & $5\%=2.4$ b  & --  & -- \\
Set size       & $95\%=6.7$ b & $30.4$ b & $7.5$ b \\
\textbf{Total} & \textbf{9.9 bits/obj} & \textbf{30.4 bits/obj} & \textbf{8.3 bits/obj} \\
\bottomrule
\end{tabular}
\label{tab:mem_overhead}
\vspace{-0.1in}
\end{table}

\section{Discussion}
\label{sec:discussion}
In this section, we discuss the limitations and applicability of \ZNSKVCache, with a focus on device compatibility, PBFG-based index design trade-off, and limitations of hotness tracking.

\textbf{Device compatibility.} 
\ZNSKVCache~ does not place special requirements on SSD or NVMe interfaces, making it compatible with both conventional SSDs and advanced ZNS or FDP SSDs. 
This compatibility stems from its coarse-grained, FIFO-style write pattern, which aligns well with the sequential write characteristics of advanced SSDs. 
When deployed on different hardware, \ZNSKVCache~ adapts by adjusting the mapping between logical units (SGs) and physical erase units. 
For instance, on large-zone ZNS SSDs, such as Western Digital ZN540 \cite{ZN540} and DapuStor J5500Z \cite{dapustor}, the configuration can follow the setup in this paper. On small-zone ZNS SSDs (e.g., Samsung PM1731a \cite{samsung_PM1731a}), where each zone is 96 MB, an SG is composed of multiple zones \cite{bae2022you, min2023ezns}. 
On FDP SSDs from Samsung with 8 GB reclaim units \cite{samsungFDPintroduction}, multiple SGs can be grouped to fill a reclaim unit.

\textbf{Balancing approximate index accuracy and read amplification.} 
In \ZNSKVCache, higher index accuracy doesn't always reduce read amplification and often increases memory overhead. 
\ZNSKVCache's read amplification arises from false positives when accessing objects, and retrieving Set-level BFs from flash. 
A more accurate index reduces the first factor, but tends to worsen the second. This is because higher accuracy comes with larger BF, which expands the index pool size. Therefore, the on-flash index group may become more scattered, leading to more flash reads. To concretely analyze this trade-off, we model the optimal accuracy for the approximate indexing used in this paper  (i.e., PBFG, detailed in Appendix \ref{appx:modeling}). Note that this model is specific to PBFG; other instantiations  may yield different optimal-accuracy models.

\textbf{Limitations of hybrid hotness tracking.}
In \ZNSKVCache, the index group aggregates Set-level BFs for a group of objects. When object hotness is estimated based on the index cache and a low-bits access counter, some cold objects may be mistakenly kept because they are part of a group marked as hot. This ``free-riding'' may reduce the accuracy of eviction. 

\section{Related Work}
\textit{\textbf{Flash-based cache.}} Existing flash-based cache architectures fall into three categories. Log-structured caches reduce WA via batched writes but incur high memory overhead for the index table \cite{fleisenman2019flashield}. Set-associative caches \cite{berg2020cachelib} minimize memory cost but suffer severe WA for tiny objects. Hierarchical designs (e.g., Kangaroo {\cite{mcallister2021kangaroo}} and FairyWREN {\cite{mcallister2024fairywren}}) balance these trade-offs, yet still exhibit high ALWA. \ZNSKVCache~ is a new architecture that integrates set-associative layout into log-style batching with reduced hash ranges to achieve near-ideal WA, and designs an on-demand PBFG indexing to maintain a low memory overhead.

\textbf{\textit{Write amplification optimization}}. Existing optimizations span multiple layers. Application-level methods include marking data hotness and merging small random writes into sequential writes through log-structured style writes \cite{f2fs}, and proactively cleaning expired data to reduce invalid writes at the source \cite{balmau2019silk}. Storage system-level strategies involve cold-hot data separation \cite{f2fs}, streamlined storage architecture \cite{vinccon2018NoFTL-KV}, optimized LSM-tree compaction strategy \cite{yao2020matrixkv}, and GC policy tuning \cite{mcallister2024fairywren}. SSD-internal improvements cover FTL hybrid mapping optimization \cite{gupta2009dftl}, prediction-based GC strategy \cite{yang2019reducingWA-byPrediction}, and multi-stream SSD supporting classified data storage to reduce valid data copying \cite{yang2017autostream}. Hardware collaboration leverages host-device synergy and NVM integration to reduce flushing frequency \cite{li2021ioda, yao2020matrixkv}.

\section{Conclusion}
This work provides a deep study of tiny objects cache systems, with a focus on the SOTA solution, FW. 
Through both theoretical analysis and practical evaluation, we identify that FW’s ALWA remains excessively high. 
To address this limitation, we propose \ZNSKVCache, a novel cache architecture that increases hash collision probability to improve per-set fill rates, thereby decreasing the ALWA. 
Additionally, \ZNSKVCache~ incorporates complementary techniques to simultaneously achieve high memory efficiency and maintain low miss ratios.

\begin{acks}
We thank the anonymous reviewers for their valuable suggestions. 
The work is supported by the National Natural Science Foundation of China (Grant No. 62572409, U22B2023, 62472061), National Key R\&D Program of China (Grant No. 2024YFB4505201, 2024YFB4504400), Chongqing Natural Science Foundation (Grant No. CSTB2025NSCQ-LZX0139), Fundamental Research Funds for the Central Universities (Grant No. 20720250166, 20720250167) and Xiaomi Young Talents Program. 
Congming Gao is the corresponding author.
\end{acks}

\normalem
\bibliographystyle{unsrt}
\balance
\bibliography{references}

\appendix
\section{Appendix: Modeling accuracy and amplification trade-off in PBFG.}\label{appx:modeling}
To concretely analyze the trade-off between index accuracy and read amplification discussed in the main body, we model the expected flash access cost of the PBFG-based approximate indexing mechanism used in this paper.

Assume an SG pool of size $N$.
Each set occupies one flash page of size $w$.
With an average object size $s$, each set contains $\frac{w}{s}$ objects.
PBFG uses bloom filters with false positive rate $x$ and a per-object memory cost of $o$, resulting in a set-level bloom filter size of $(w/s)\cdot o$ bits.
Let $n$ be the number of set-level bloom filters per flash page. Then
\[ 
n = \frac{w}{(w/s)\cdot o} = \frac{s}{o} 
\]
As a result, retrieving PBFGs for $N$ SGs requires reading approximately $N/n$ flash pages.

During a lookup, one object is always accessed, and BF false positives may additionally trigger accesses to objects in other SGs.
In expectation, this results in accessing $1 + (N-1)x$ objects from flash.
Combining these two components, the total expected flash access cost can be approximated as:
\begin{equation}
\label{eq:pbfg_cost}
\min \left( \frac{N\cdot o}{s} + 1 + (N-1)x \right).
\end{equation}

When $N$ and $s$ are fixed,~\eqref{eq:pbfg_cost} can be rewritten as
\begin{equation}
\label{eq:pbfg_cost_linear}
\min \left( C_1 \cdot o + C_2 \cdot x + C_0 \right),
\end{equation}
where $C_0$, $C_1$, and $C_2$ are constants.
In bloom filters, the memory overhead $o$ increases monotonically as the false positive rate $x$ (i.e., approximate index accuracy) decreases. As a result, there exists an optimal BF configuration that minimizes the overall flash reads.

\textbf{Instantiation with evaluation parameters.}
Under our evaluation settings, the SG pool contains $N=350$ SGs.
In the worst case, a lookup in \ZNSKVCache reads PBFGs from 7 flash pages, followed by flash accesses to candidate objects.
With a BF false positive rate of $0.1\%$, the expected number of object-related flash accesses is approximately $1 + 0.35$.
Reducing the BF false positive rate to $0.01\%$ increases the PBFG retrieval cost to 9 flash pages, while reducing object-related flash accesses to approximately $1 + 0.03$.
As a result, the expected total number of flash reads increases from approximately $7 + 1 + 0.35$ to $9 + 1 + 0.03$.
The reduction in false positives may increase read amplification, failing to offset the increased PBFG retrieval cost.

\textbf{Impact of scaling flash capacity.}
Increasing flash capacity increases the number of SGs $N$, which may appear to increase the objective in~\eqref{eq:pbfg_cost}.
In practice, this effect can be mitigated by partitioning the flash device into multiple independent regions, each managed by a separate flash cache instance.
With partitioning, the effective SG pool size per instance remains bounded, limiting the impact of flash capacity on read amplification.

\section*{Artifact Appendix}

\subsection{Abstract}

This artifact implements \ZNSKVCache, a flash-based cache architecture for tiny objects, as a C++ module built on Facebook's CacheLib engine. It contains all components required to reproduce the key results of the paper ``Nemo: A Low-Write-Amplification Cache for Tiny Objects on Log-Structured Flash Devices.'' The artifact includes: (1) the Nemo implementation and a modified FairyWREN baseline in C++ and (2) scripts for automated benchmarking and measurement. Using this artifact, reviewers can remotely replay the Twitter trace on a real ZNS SSD using \ZNSKVCache and baseline schemes, and validate \ZNSKVCache's improvements in write amplification, median and tail latency, and hit ratio.

\subsection{Artifact check-list (meta-information)}

\begin{itemize} [leftmargin=14pt, itemsep=0pt, topsep=2pt]
  \item {\bf Compilation:} C++17.
  \item {\bf Run-time environment:} Linux. A recent kernel with Zoned Namespace support is required, for example Ubuntu~22.04 with Linux kernel~$\ge$~5.15.
  \item {\bf Hardware:} Specialized hardware required (ZNS SSD).
  \item {\bf Metrics:} Write amplification, read latency, hit ratio.
  \item {\bf Output:} Several binaries and log files.
  \item {\bf How much time is needed to prepare the workflow (approximately)?:} One-click compilation and setup. Approximately 30 minutes.
  \item {\bf How much time is needed to complete experiments (approximately)?:} One to two weeks. Experiments can run unattended after the scripts are launched.
  \item {\bf Publicly available?:} Yes.
  \item {\bf Code licenses:} Apache License.
  \item {\bf Data licenses:} CC-BY-4.0 license.
\end{itemize}

\subsection{Description}

\subsubsection{How to access}

GitHub: (\href{https://github.com/XMU-DISCLab/Cachelib-Nemo}{https://github.com/XMU-DISCLab/Cachelib-Nemo}) or Zenodo: (\href{https://doi.org/10.5281/zenodo.18332674}{https://doi.org/10.5281/} \href{https://doi.org/10.5281/zenodo.18332674}{zenodo.18332674}).

\subsubsection{Hardware dependencies}
The system requires more than 32~GB of memory. The evaluation machine is equipped with a 14~TB Western Digital ZN540 ZNS SSD. If a ZNS SSD is not available locally, the experiments cannot be fully reproduced on alternative hardware. 

\subsection{Installation and Testing} 

\subsubsection{Installation:} Installation instructions are provided in the GitHub repository README.

\subsubsection{Basic Test:} Basic testing instructions are also described in the repository README. If Nemo and FairyWREN runs correctly, three types of log files will be generated under the \texttt{cachelib/log} directory: ``\texttt{summary.log}'', ``\texttt{run.log}'', and ``\texttt{progress.log}''.

\subsection{Evaluation and expected results}

The artifact enables reproduction of the following results:

\begin{itemize} [leftmargin=14pt, itemsep=0pt, topsep=2pt]
  \item \textbf{Write amplification} (Figure~11). Comparison of WA between \ZNSKVCache and FairyWREN.
  \item \textbf{Write amplification trends} (Figure~13). WA evolution of \ZNSKVCache and FairyWREN as the trace progresses.
  \item \textbf{Read latency} (Figure~14). Median, p99, and p9999 read latency for \ZNSKVCache and FairyWREN.
  \item \textbf{Miss ratio} (Figure~15). Hit and miss ratio trends for \ZNSKVCache and FairyWREN over the trace.
\end{itemize}

\end{document}